\documentclass[12pt]{article}
\usepackage{latexsym}
\begin{document}

\title{{\bf How to Describe Photons as (3+1)-Solitons?}}
\author{{\bf S.Donev$^1$, D.Trifonov$^2$}\\ Institute for Nuclear Research
and Nuclear Energy,\\ Bulg.Acad.Sci., 1784 Sofia, blvd.Tzarigradsko chaussee
72\\ Bulgaria\\ e-mails: $^1$sdonev@inrne.bas.bg,\ $^2$dtrif@inrne.bas.bg}

\date{}

\maketitle

\begin{abstract}
This paper aims to present the pure field part of the newly developed
nonlinear {\it Extended Electrodynamics} [1]-[3] in non-relativistic terms,
i.e. in terms of the electric and magnetic vector fields (${\mathbf
E},{\mathbf B}$), and to give explicitly those (3+1)-soliton solutions of the
new equations which have the integral properties of photons.  The set of
solutions to the new equations contains all solutions to Maxwell's equations
as a subclass, as well as, new solutions, called nonlinear. The important
characteristics {\it scale factor}, {\it amplitude function}, and {\it phase
function} of a nonlinear solution are defined in a coordinate free way and
effectively used.  The nonlinear solutions are identified through the
non-zero values of two appropriately defined vector fields $\vec{\cal F}$ and
$\vec{\cal M}$, as well as, through the finite values of the corresponding
scale factors. The intrinsic angular momentum (spin) is also defined.  A
limited superposition principle (interference of nonlinear solutions),
yielding the well known classical {\it coherence} conditions, is found to
exist.

\end{abstract}

\newpage

\section{Introduction}

The 19th century physics, due mainly to Faraday and Maxwell, created the
theoretical concept of {\it electromagnetic field}, i.e. extended (in fact,
infinite) object, having dynamical structure. The concepts of {\it flux of a
vector field through a 2-dimensional surface} and {\it circulation of a
vector field along a closed curve} were coined and used extensively.
Maxwell's equations in their integral form establish where the time-changes
of the fluxes of the electric and magnetic fields go to, or come from, in
both cases of a closed 2-surface and a 2-surface with a boundary. We note
that these fluxes are specific to the continuous character of the physical
object under consideration and it is important also to note that Maxwell's
field equations have not the sense of direct energy-momentum balance
relations as the Newton's law $\dot{\mathbf p}={\mathbf F}$ has.
Nevertheless, they are consistent with energy-momentum conservation, as is
well known, the corresponding local energy-momentum quantities are quadratic
functions of the electric and magnetic vectors.

Although very useful for considerations in finite regions with boundary
conditions, the pure field Maxwell's equations have time-dependent vacuum
solutions (in the whole space) that give inadequate models of the
real fields.  As a rule, if these solutions are time-stable, they occupy the
whole 3-space or an infinite subregion of it, and they do not go to zero
at infinity, hence, they carry {\it infinite} energy and momentum.  As an
example we recall the plane wave solution, given by the electric and magnetic
fields of the form (in a specally chosen coordinate system)
\[
{\mathbf E}=\Bigl\{ u(ct+\varepsilon z),
p(ct+\varepsilon z), 0\Bigr\} ;\
{\mathbf B}=\Bigl\{\varepsilon p(ct+\varepsilon z),
-\varepsilon u(ct+\varepsilon z), 0\Bigr\} ,\ \varepsilon=\pm1,
\]
where $u$ and $p$ are arbitrary differentiable functions. Even if $u$ and $p$
are soliton-like with respect to the coordinate $z$, they do not depend on
the other two spatial coordinates $(x,y)$. Hence, the solution occupies the
whole ${\cal R}^3$, or its infinite subregion, and clearly it carries
infinite integral energy (we use Gauss units)
\[
W=\frac{1}{4\pi}\int_{{\cal R}^3}\frac{{\bf E}^2+{\bf B}^2}{2}dxdydz=
\frac{1}{4\pi}\int_{{\cal R}^3}(u^2+p^2)dxdydz=\infty.
\]
In particular, the popular harmonic plane wave
$$
u=U_o{\rm cos}(\omega t\pm k_z.z),\ p=P_o {\rm sin}(\omega t\pm k_z.z),\
c^2 k_z^2=\omega^2,\ U_o=const,\ P_o=const,
$$
clearly occupies the whole 3-space, carries infinite energy
$$
W=\frac{1}{4\pi}\int_{{\cal R}^3}(U_o+P_o)dxdydz =\infty
$$
and, therefore, could hardly be an adequate model of a really created field.

On the other hand, according to Cauchy's theorem for the wave equation
$\Box U =0$ (which
is necessarily satisfied by the components of ${\mathbf E}$ and ${\mathbf B}$
in the pure field case), every finite (and smooth enough) initial field
configuration is strongly time-unstable [4]: the initial condition blows up
radially and goes to infinity (see the next section).  Hence, Maxwell's
equations {\it cannot describe finite and time-stable localized fields}. The
contradictions between theory and experiment that became clear at the end of
the last century were a challenge to theoretical physics.  Planck and
Einstein created the notion of {\it elementary field quanta}, named later by
Lewis [5] the {\it photon}. The concept of {\it photon} proved to be very
seminal and has been widely used in the 20th century physics.  However, even
now, after almost a century, we still do not have a complete and satisfactory
self-consistent theory of single photons. It worths recalling at this place
Einstein's remarks of unsatisfaction concerning the linear character of
the vacuum Maxwell theory which makes it not able to describe the
microstructure of radiation [6].  Along this line we may also note here some
other results and oppinions [7]-[8].

In this paper we consider {\it single photons} as (3+1)-dimensional {\it
extended finite wave-like objects}, or (3+1) {\it solitary waves}, moving as
a whole in a consistent {\it translational-rotational} manner with the speed
of light. Their integral characteristics like {\it frequency, period} and
{\it spin} are considered to be intrinsically related to their {\it periodic
dynamical structure}, and the most important integral quantity characterizing
the single photon seems to be its {\it intrinsic elementary action}, this
being put equal to Planck's constant $h$.  The usually made in
quantum theory assumption for a point-like character of photons
we consider as {\it inadequate}, it can never give a satisfactory explanation
of where the characteristic {\it frequency} of a {\it free} photon comes
from.  Indeed, the point approximation means {\it structurelessness}, so, the
frequency may appear only if an external force acts on the point particle,
but this means that the particle {\it is not free}, hence, its velocity
must not be constant and its spin momentum {\it is not an intrinsic}
characteristic, which contradicts the usual understanding of Planck's formula
$E=h\nu$.  In other words, all objects that obey the Planck's formula
$E=h\nu$ do {\it not} admit point approximation, hence, the solitary wave
view, considered as a first step, seems much more natural.  {\it Extended
Electrodynamics} (EED) [1]-[3] was built mainly to meet this solitary wave
view, i.e. to give a consistent field description of single photons, and this
paper gives a completely non-relativistic approach to the pure field part of
EED.  This gives new insights into the problem.

Our assumption that single photons are objects of a finite solitary wave
nature has as its mathematical representation the soliton concept. In
accordance with this concept the components of the corresponding ${\mathbf
E}$ and ${\mathbf B}$ at every moment $t$ have to be smooth
nonsingular functions and different from zero only inside a finite
3-dimensional region $\Omega_t \subset  {\cal R}^3$.  Moreover, a
$t$-periodic process of constant frequency has to accompany the photon's
translational motion as a whole.  Hence, we have to be ready to meet all
difficulties coming from the unavoidable requirements for using nonlinear
partial differential equations having finite with respect to the spatial
variables $(x,y,z)$ solutions. This goes along with the Einstein's view that
"the whole theory must be based on partial differential equations and their
singularity-free solutions"[9].

\section{The new equations}
We recall first Maxwell's equations in the pure field case:
\begin{equation}
rot{\mathbf B}-\frac{\partial {\mathbf E}}{\partial \xi}=0,\ \ \
div{\mathbf E}=0,                                             
\end{equation}

\begin{equation}
rot{\mathbf E}+\frac{\partial {\mathbf B}}{\partial \xi}=0,\ \ \
div{\mathbf B}=0,                                              
\end{equation}
where $\xi$ denotes the product $ct$, $c$ is the velocity of light in
vacuum and $t$ is the time variable. From these equations we get the well
known Poynting relation
\begin{equation}
\frac{\partial}{\partial \xi}\frac{{\mathbf E}^2 +{\mathbf B}^2}{8\pi}=
-\frac{1}{4\pi}div({\mathbf E}\times{\mathbf B}).                   
\end{equation}

We explain now why Maxwell's equations (1)-(2) have no (3+1) soliton-like
solutions. As we know from textbooks on Classical
Electrodynamics (CED), (e.g. [10])
from (1) and (2) it follows that every component $U$
of ${\mathbf E}$ and ${\mathbf B}$ {\it necessarily} satisfies the wave
equation
\begin{equation}
\Box U\equiv U_{tt}-c^2\left[U_{xx}+U_{yy}+U_{zz}\right]=0.        
\end{equation}
We are interested in the behavior of $U$ at $t>0$, if at $t=0$ the function
$U$ satisfies the initial conditions
\[
U|_{t=0}=f(x,y,z),\quad \frac{\partial U}{\partial t}\biggl|_{t=0}=F(x,y,z) ,
\]
where the functions $f(x,y,z)$ and $F(x,y,z)$ are finite, i.e. they are
different from zero in some finite region $\Omega_o \subset{\cal R}^3$ and
have no singularities. Besides, we assume also that $f$ is continuously
differentiable up to third order, and $F$ is continuously differentiable up
to the second order. Under these conditions Poisson proved (about 1818) that
a unique solution $U(x,y,z;t)$ of the wave equation is defined, and it is
expressed by the initial conditions $f$ and $F$ through the following
formula (a good explanation is given in [4]):
\begin{equation}
U(x,y,z,t)=\frac{1}{4\pi c}
\left\{\frac{\partial}{\partial t}\Biggl[\int_
{S_{ct}}\frac{f(P)}{r}d\sigma_r \Biggr]+\int_{S_{ct}}\frac{F(P)}{r}
d\sigma_r \right\},                                             
\end{equation}
where $P$ is a point on the sphere $S$ of radius $r=ct$, centered at the
point $(x,y,z)$, and $d\sigma_r$ is the surface element on $S_{r=ct}$.

The above formula (5) shows the following. In order to get the solution at
the moment $t>0$ at the point $(x,y,z)$, being at an arbitrary position with
respect to the region $\Omega_o$, where the initial condition, defined by the
two functions $f$ and $F$, is concentrated, we have to integrate $f/r$ and
$F/r$  over a sphere $S_{r=ct}$, centered at $(x,y,z)$ and having a radius of
$r=ct$, and then to form the expression (5).  Clearly, the solution will be
different from zero at the moment $t>0$ only if the sphere $S_{r=ct}$ crosses
the region $\Omega_o$ at this moment.  Consequently, if $r_1=ct_1$ is the
shortest distance from $(x,y,z)$ to $\Omega_o$, and $r_2=ct_2$ is the longest
distance from $(x,y,z)$ to $\Omega_o$, then the solution at $(x,y,z)$ will be
different from zero only inside the interval $(t_1,t_2)$.

From the point of view of an external observer this means the following. The
initially concentrated in the region $\Omega_o$ perturbation begins to
expand, it comes to an arbitrary point $(x,y,z)$ at the moment $t_1>0$, makes
it "vibrate" ( i.e. our devices show the availability of field at this point)
during the time interval $\Delta t=t_2-t_1$, after this the disturbed point
goes back to its initial state and our devices find no more field there.
Through every point out of $\Omega_o$ there will pass a wave, and its
forefront reaches the point $(x,y,z)$ at some moment $t_1$ while its backfront
leaves the same point at the moment $t_2>t_1$.  Roughly speaking, the initial
condition "blows up radially" and goes to infinity with the velocity of
light.

This rigorous mathematical result shows that {\it every} finite initial
condition induces strongly time-unstable free finite time-dependant solution
of Maxwell's equations in vacuum, so these equations {\it have no} finite and
smooth enough, i.e. nonsingular, time-dependent solutions, which could be
used as models of real photons, as viewed by us.

Hence, if we want to describe 3-dimensional time-dependent soliton-like
electromagnetic formations (or configurations) it is {\it necessary} to leave
off Maxwell's equations and to look for new equations for ${\mathbf E}$ and
${\mathbf B}$.

On the other hand we know that Maxwell's theory is widely used in almost all
natural sciences and electrical engineering, so, it does not seem reasonable
to leave it off entirely and to look for a completely new theory. Moreover,
in all energy computations for finite volumes, it gives very good results.
This suggests, at this stage, to look for some extension of the theory, i.e.
to extend in a nonlinear way the equations, keeping all solutions to
Maxwell's equations as solutions to the new equations and keeping the
energy-momentum relations of Maxwell's theory as relations of the new theory.
In the same time we must incorporate new solutions with corresponding to our
purpose properties.

The road we are going to follow in searching for the appropriate
nonlinearization of (1)-(2) is suggested mainly by two ideas: the idea of
{\it local energy-momentum conservation}, and the idea of {\it
invariance} of Maxwell's equations (1)-(2) with respect to the
transformation
\begin{equation}     
({\mathbf E},{\mathbf B})\rightarrow
(-{\mathbf B},{\mathbf E}).
\end{equation}

We begin by recalling the second Newton's law in mechanics: $\dot{\mathbf
p}$=${\mathbf F}$, its true sense is {\it local momentum balance}, i.e. the
momentum gained by the particle is lost by the external field. So, if the
field is absent: ${\bf F}=0$, then the particle will not lose, or gain,
momentum, and we get the evolution equation for a free particle: $\dot{\bf
p}=m\dot{\bf v}=0$.  On the other hand, if the particle is absent then the
field will not change its momentum and we get the equation ${\mathbf F}=0$.
Usually, ${\bf F}$ depends on the characteristics of the particle, as well
as, on the characteristics of the field. In order to make possible the
interpretation of the  relation ${\bf F}=0$ as a pure-field equation we have
to express, if possible, this ${\mathbf F}$ in terms of the corresponding
 field functions (and their derivatives) only.  Fortunately, this can be done
in electrodynamics, Maxwell's eqiations in presence of sources make it
possible.  In this way we shall obtain explicitly in terms of the field
functions and their derivatives one of the possible expressions describing
how much momentum the field is potentially able to transfer locally to
another physical system in case of its presence at the corresponding point
and its ability to absorb this momentum, i.e. to interact with the
field.  Of course, the field may have various such momentum (and
corresponding energy) exchanging abilities.  We'd like to note at this moment
that the following considerations (up to equation (7)) have a suggestive
nature only, they do not prove what we are going to assume finally as the new
equations.

In order to carry out the above idea  we recall first that if the other
system consists of charged particles, as it is in Maxwell theory, the
corresponding force is the well known Lorentz' force, acting on a particle of
electric charge $e$:  ${\mathbf F}=e{\mathbf E}+\frac ec({\mathbf v}\times
{\mathbf B})$.  In case of a continuous distribution of particles with charge
density of $\rho$ and current ${\bf j}=\rho{\bf v}$ the Lorentz' force is
\[
{\mathbf F}=\rho{\mathbf E}+\frac{\rho}{c}({\mathbf v}\times {\mathbf B})=
\rho{\mathbf E}+\frac 1c({\mathbf j}\times {\mathbf B}).
\]
The corresponding Maxwell's equations with non-zero charge distribution
$\rho$ and current ${\mathbf j}=\rho{\bf v}$ in this case read:
\[
rot{\mathbf B}-\frac{\partial {\mathbf E}}{\partial \xi} =
\frac{4\pi}{c}{\mathbf j},
\ \ div{\mathbf E}=4\pi\rho.
\]
These last equations make possible to substitute ${\mathbf j}$ and
$\rho$ into the above given Lorentz' force. Having this done
we put ${\mathbf F}=0$ and then we forget about the character of the
"other system" (charged particles). We interpret now ${\mathbf F}$, so
obtained and expressed through the field functions and their derivatives
only, as a definite quantity of momentum which the field is {\it potentially
able} to transfer (locally) to any (continuous) physical system that is able
to absorb it.  This defines quantitatively one of the field's momentum
exchanging abilities.

The above suggestive considerations make us assume our first extended
equation:
\begin{equation}
\left(rot{\mathbf B}-\frac{\partial {\mathbf E}}{\partial \xi}\right)
\times {\mathbf B}+{\mathbf E}div{\mathbf E}=0 .      
\end{equation}
This vector equation (7) extends Maxwell's pure field equations (1) in the
sense that (7) implies no more (1), i.e. (7) may have solutions which do not
satisfy (1) and these new solutions, at least some of them, are considered as
{\it admissible} from physical point of view, i.e. describing some physical
reality.  The physical sense of (7) is quite clear:  {\it no field momentum
is lost in this way}. At the same time relation (7) describes some internal,
i.e. between ${\bf E}$ and ${\bf B}$, redistribution of the field
energy-momentum during the field's time evolution. The nonlinearity of (7)
is also obvious.  We'd like to stress once again the suggestive character of
the obove considerations, equation (7) is an assumption, we {\it do not
consider it as a consequence of Maxwell equations with non-zero current}, on
the contrary, we consider it as a {\it pure field} equation.

Now we look for another momentum exchanging ability of the field, different
from the one given by the left-hand side of (7), and directed in general to
new physical systems. In order to come to
such an expression we make use of the above mentioned $({\mathbf E},{\mathbf
B})\rightarrow (-{\mathbf B},{\mathbf E})$ invariance of the pure field
Maxwell's equations, known as electro-magnetic duality.  This invariance is
valid also for the energy $w$ and momentum $\vec S$ (i.e. Poynting's vector)
densities:
$$
w=\frac{{\mathbf E}^2+{\mathbf B}^2}{8\pi}\rightarrow
\frac{{\mathbf B}^2+{\mathbf E}^2}{8\pi}, \ \
\vec S=\frac{c}{4\pi}{\mathbf E}\times {\mathbf
B}\rightarrow -\frac{c}{4\pi}{\mathbf B}\times {\mathbf E}=
\frac{c}{4\pi}{\mathbf E}\times {\mathbf B}.
$$
Moreover, the basic energy-momentum balance relation of Poynting (3), is also
invariant with respect to this transformation (6). This suggests that,
transforming (7), i.e. replacing in (7) ${\mathbf E}$ by $-{\mathbf B}$ and
${\mathbf B}$ by ${\mathbf E}$, we should obtain also a true and valuable
relation, since (7) describes now intra-field local momentum balance.  In this
way we obtain our second vector equation:
\begin{equation}
\left(rot{\mathbf E}+\frac{\partial {\mathbf B}}{\partial \xi}\right) \times
{\mathbf E}+{\mathbf B}div{\mathbf B}=0.           
\end{equation}
The left-hand side of (8) defines explicitly another momentum exchanging
ability of the field, and relation (8) defines another way of internal field
energy-momentum redistribution with time. Note that (8) is obtained from (7)
in the same way as (2) is obtained from (1).

We complete this process of extension of Maxwell's equations
by adding two new invariant with respect to the same transformation (6)
equations, which also have the physical sense of intra-field
local energy-momentum balance:
\begin{equation}
\left(rot{\mathbf E}+\frac{\partial {\mathbf
B}}{\partial \xi}\right) \times {\mathbf B}+ \left(rot{\mathbf
B}-\frac{\partial {\mathbf E}}{\partial \xi}\right) \times {\mathbf
E}-{\mathbf E}div{\mathbf B}
-{\mathbf B}div{\mathbf E}=0 ,             
\end{equation}
\begin{equation}
{\mathbf B}.\left(rot{\mathbf B}-
\frac{\partial {\mathbf E}}{\partial \xi}\right)-
{\mathbf E}.\left(rot{\mathbf E}
+\frac{\partial {\mathbf B}}{\partial \xi}\right)=0 .     
\end{equation}
\noindent
{\bf Remark}: In the relativistic formulation of EED [3] the two relations
(9) and (10) have a natural interpretation of energy-momentum transfers
between $F$ and $*F$, where $F$ is the conventional electromagnetic field
tensor constructed by means of ${\mathbf E}$ and ${\mathbf B}$, and $*F$ is
its dual (constructed similarly by means of $-{\mathbf B}$ and ${\mathbf E}$).
This interpretation says that these transfers are mutual:
$F\rightleftharpoons *F$, and always in equal quantities. In other words, the
relativistic formalism considers the couples (${\mathbf E},{\mathbf B}$) and
($-{\mathbf B},{\mathbf E}$), or $F$ and $*F$, as two componenents of a new
more general mathematical object [2]-[3].

\vskip 0.5cm

Note that under the transformation (6) equations (7) and (8) transform into
each other, while equations (9) and (10) are kept the same (up to a sign of
the left-hand side).

Equations (7)-(10) constitute our new system of equations for the
electromagnetic field in vacuum.  Obviously, they do not introduce new
parameters, they are non-linear and all solutions to Maxwell's vacuum
equations (1)-(2) are solutions to the new equations (7)-(10).  We are going
now to study those new solutions to (7)-(10), which satisfy the conditions
\begin{equation}
rot{\mathbf E}+\frac{\partial {\mathbf B}}{\partial \xi}\neq 0,
\ \ rot{\mathbf B}-\frac{\partial {\mathbf E}}{\partial \xi}\neq 0,
\ \ div{\mathbf E} \neq 0,\ \ div{\mathbf B}\neq 0.                 
\end{equation}
For further convenience, all solutions to (7)-(10), which satisfy (11), will
be called {\it nonlinear}.

\section{Properties of the Nonlinear Solutions}
The first two, almost obvious, properties of the nonlinear solutions
follow directly from (7) and (8) and are given by the relations
\begin{equation}
{\mathbf E}.{\mathbf B}=0,                                     
\end{equation}
\begin{equation}
{\mathbf B}.\left(rot{\mathbf E}+
\frac{\partial {\mathbf B}}{\partial \xi}\right)=0,\ \
{\mathbf E}.\left(rot{\mathbf B}-                                
\frac{\partial {\mathbf E}}{\partial \xi}\right)=0.
\end{equation}
Relation (12) says that the electric and magnetic vectors of every nonlinear
solution are {\it necessarily orthogonal} to each other at every point,
the algebraic property that Maxwell's (linear) plane wave solution has. From
(13) it follows that the Poynting's relation (3) is true for all nonlinear
solutions, and this justifies the usage of all energy-momentum quantities and
relations from Maxwell's theory in the set of nonlinear solutions of the new
equations. We can consider the left-hand sides of relations (13) as the
energy quantities which the field is potentially able to transfer to some
other physical object. We note also the obvious invariance of (12)-(13) with
respect to transformation (6).

We are going to show now that all nonlinear solutions satisfy the relation
\begin{equation}
{\mathbf E}^2={\mathbf B}^2 .                                    
\end{equation}
In order to prove (14) let's take the scalar product of equation (8) from the
left by ${\mathbf B}$.  We obtain

\[
{\mathbf B}.\Biggl\{\left(rot{\mathbf E}+
\frac{\partial {\mathbf B}}{\partial \xi}\right)
\times {\mathbf E}\Biggr\}+{\mathbf B}^2div{\mathbf B}=0. \ \ \ \ \ (*)
\]
Now, multiplying (9) from the left by ${\mathbf E}$ and having in view (12),
we obtain
\[
{\mathbf E}.\Biggl\{\left(rot{\mathbf E}+
\frac{\partial {\mathbf B}}{\partial \xi}\right)
\times {\mathbf B}\Biggr\}-{\mathbf E}^2div{\mathbf B}=0.
\]
This last relation is equivalent to
\[
-{\mathbf B}.\Biggl\{\left(rot{\mathbf E}+
\frac{\partial {\mathbf B}}{\partial \xi}\right)
\times {\mathbf E} \Biggr\}-
{\mathbf E}^2div{\mathbf B}=0.\ \ \ \\ (**)
\]
Now, summing up $(*)$ and $(**)$, in view of $div{\mathbf B}\neq 0$,
we obtain (14).

Relation (14) is also true for the linear plane electromagnetic wave. It
requires for all nonlinear solutions a permanent equality of the energy
densities carried by the electric and magnetic fields, although permanent
mutual energy-momentum flows run between ${\bf E}$ and ${\bf B}$, which means
that these two flows are always in equal quantities.

We introduce now the following two vector fields:
\begin{equation}
\vec{\cal F}=rot{\mathbf E}+\frac{\partial {\mathbf B}}{\partial \xi}+
\frac{{\mathbf E}\times
{\mathbf B}}{|{\mathbf E}\times {\mathbf B}|}div{\mathbf B},   
\end{equation}
\begin{equation}
\vec{\cal M}=rot{\mathbf B}-\frac{\partial {\mathbf E}}{\partial \xi}-
\frac{{\mathbf E}\times
{\mathbf B}}{|{\mathbf E}\times {\mathbf B}|}div{\mathbf E}.    
\end{equation}

It is obvious that on the solutions of Maxwell's equations (1)-(2) $\vec{\cal
F}$ and $\vec{\cal M}$ are equal to zero. Note also that under the
transformation (6) we get $\vec{\cal F}\rightarrow -\vec{\cal M}$ and
$\vec{\cal M}\rightarrow \vec{\cal F}$.  We shall show now that on the
non-zero nonlinear solutions of our equations (7)-(10) $\vec{\cal F}$ and
$\vec{\cal M}$ are colinear to ${\bf E}$ and ${\bf B}$ respectively.
Indeed, consider the products $\vec{\cal F}\times {\mathbf E}$ and
$\vec{\cal M}\times{\mathbf B}$. Since
$$
({\mathbf E}\times {\mathbf
B})\times {\mathbf E}= -{\mathbf E}\times ({\mathbf E}\times {\mathbf B})=
-[{\mathbf E}({\mathbf E}.{\mathbf B})-
{\mathbf B}({\mathbf E}.{\mathbf E})]={\mathbf B}|{\mathbf E}|^2
$$
and $|{\mathbf E}\times {\mathbf B}|={\mathbf E}^2={\mathbf B}^2$,
we obtain (see eqn.(8))
$$
\vec{\cal F}\times {\mathbf E}=
\left(rot{\mathbf E}+\frac{\partial {\mathbf B}}{\partial \xi}\right)
\times {\mathbf E}+{\mathbf B}div{\mathbf B}=0.
$$
In the same way we get $\vec{\cal M}\times {\mathbf B}=0$. Hence, we
can write the relations
\begin{equation}
\vec{\cal F}=f_1.{\mathbf E},\ \ \vec{\cal M}=f_2.{\mathbf B},        
\end{equation}
where $f_1$ and $f_2$ are two functions, and further we consider the
interesting cases $f_1\neq 0,\infty;\linebreak \ f_2\neq 0,\infty$.
Note that the physical dimension of $f_1$ and $f_2$ is the
reciprocal to the dimension of coordinates, i.e.
$[f_1]=[f_2]= [length]^{-1}$. Note also, that ${\cal F}$ and ${\cal M}$ are
mutually orthogonal.

We shall prove now that $f_1=f_2$. In fact, making use of the same formula
for the double vector product, used above, we easily obtain (see eqn.(9))
\[
\vec{\cal F}\times {\mathbf B}+\vec{\cal M}\times {\mathbf E}=
\]
\[
=\left(rot{\mathbf E}+\frac{\partial {\mathbf B}}{\partial \xi}\right)
\times {\mathbf B}+
\left(rot{\mathbf B}-\frac{\partial {\mathbf E}}{\partial \xi}\right)
\times {\mathbf E}-{\mathbf E}div{\mathbf B}
-{\mathbf B}div{\mathbf E}=0 .
\]
Therefore,
\[
\vec{\cal F}\times {\mathbf B}+\vec{\cal M}\times {\mathbf E}=
\]
\[
=f_1{\mathbf E}\times {\mathbf B}+f_2{\mathbf B}\times {\mathbf E}=
(f_1-f_2){\mathbf E}\times {\mathbf B}=0.
\]
The assertion follows. Now from (14) and (17) it follows also the relation
$|\vec{\cal F}|=|\vec{\cal M}|$.
\vskip 0.5cm
{\bf Definition 1}. The quantity
\begin{equation}
L({\mathbf E},{\mathbf B})=\frac{1}{|f_1|}=\frac{1}{|f_2|}=
\frac{|{\mathbf E}|}{|\vec{\cal F}|}=        
\frac{|{\mathbf B}|}{|\vec{\cal M}|},
\end{equation}
will be called {\it the scale factor} for the nonlinear solution
$({\mathbf E},{\mathbf B})$.

Obviously, $L({\mathbf E},{\mathbf B})=L(-{\mathbf B},{\mathbf E})$, and
for all non-zero nonlinear solutions we have \linebreak
$0<L<\infty$, while for the
linear solutions $L\rightarrow \infty$.

\vskip 0.5cm

\section{Photon-Like Solutions}
As we mentioned earlier, we consider photons as finite nonsingular objects
moving translationaly along {\it straight lines} in the 3-space with the
velocity of light.  The direction of motion is assumed to be that of
$({\mathbf E}\times {\mathbf B})$.  This means that the integral curves of
$({\mathbf E}\times {\mathbf B})$ have to be straight lines.  If we choose the
coordinate system $(x,y,z)$ so that this direction of the translational
motion to coincide with the direction of the coordinate line $z$, in this
coordinate system the vector fields ${\mathbf E}$ and ${\mathbf B}$ will have
non-zero components only along $x$ and $y$:
\[
{\mathbf E}=(u,p,0);\ \ {\mathbf B}=(m,n,0),
\]
so, $({\mathbf E}\times {\mathbf B})=(0,0,un-pm)$. Now from
${\mathbf E}.{\mathbf B}=0$ and ${\mathbf E}^2={\mathbf B}^2$   it
follows $m=\varepsilon p$ and $n=-\varepsilon u$, $\varepsilon =\pm 1$.
Hence,
\[
{\mathbf E}=(u,p,0);\ \ {\mathbf B}=(\varepsilon p,-\varepsilon u,0),
\ \ {\mathbf E}\times {\mathbf B}=[0,0,-\varepsilon (u^2+p^2)],
\]
and we have to determine just the two functions $u$ and $p$.

Let's substitute these ${\mathbf E}$ and ${\mathbf B}$ into the left hand
sides of equations (7)-(10).  We obtain:
\[
\left(rot{\mathbf E}+ \frac{\partial {\mathbf B}}{\partial \xi}\right)
\times{\mathbf E}+{\mathbf B}div{\mathbf B}=
[0,0,\varepsilon p(p_\xi-\varepsilon p_z)+
\varepsilon u(u_\xi-\varepsilon u_z)];
\]
\[
\left(rot{\mathbf B}-\frac{\partial {\mathbf E}}{\partial \xi}\right)
\times{\mathbf B}+{\mathbf E}div{\mathbf E}=
[0,0,\varepsilon u(u_\xi-\varepsilon u_z)+
\varepsilon p(p_\xi-\varepsilon p_z)];
\]
\[
\left(rot{\mathbf E}+ \frac{\partial {\mathbf B}}{\partial \xi}\right)
\times{\mathbf B}=[\varepsilon u(p_x-u_y), \varepsilon p(p_x-u_y),
-u(p_\xi-\varepsilon p_z)+p(u_\xi-\varepsilon u_z)];
\]
\[
\left(rot{\mathbf B}- \frac{\partial {\mathbf E}}{\partial \xi}\right)
\times{\mathbf E}=[\varepsilon p(u_x+p_y), \varepsilon u(u_x+p_y),
-p(u_\xi-\varepsilon u_z)+u(p_\xi-\varepsilon p_z)];
\]
\[
-{\mathbf E}div{\mathbf B}=[-\varepsilon u(p_x-u_y),
-\varepsilon p(p_x-u_y),0];
\]
\[
-{\mathbf B}div{\mathbf E}=[-\varepsilon p(u_x+p_y),
\varepsilon u(u_x+p_y),0];
\]
\[
-{\mathbf E}.
\left(rot{\mathbf E}+ \frac{\partial {\mathbf B}}{\partial \xi}\right)=
-\varepsilon u(p_\xi-\varepsilon p_z)+\varepsilon p(u_\xi-\varepsilon u_z);
\]
\[
{\mathbf B}.
\left(rot{\mathbf B}- \frac{\partial {\mathbf E}}{\partial \xi}\right)=
-\varepsilon p(u_\xi-\varepsilon u_z)+\varepsilon u(p_\xi-\varepsilon p_z),
\]
where the indices of $u$ and $p$ mean the corresponding derivatives.

It is seen that equations (9)-(10) are satisfied identically, and equations
(7)-(8) reduce to only one equation, namely
\begin{equation}
u(u_\xi-\varepsilon u_z)+p(p_\xi-\varepsilon p_z)=           
\frac12\bigl[(u^2+p^2)_\xi-\varepsilon (u^2+p^2)_z\Bigr]=0.
\end{equation}
The solution to this equation is
\begin{equation}
u^2+p^2=\phi^2(x,y,\xi+\varepsilon z),                      
\end{equation}
where $\phi$ is an arbitrary differentiable function of its arguments.
This relation means that the energy density
$$
\frac{1}{8\pi}({\mathbf E}^2+{\mathbf B}^2)=
\frac{1}{4\pi}(u^2+p^2)
$$
is a running wave along the coordinate $z$.  Hence, our equations determine
the field components $u$ and $p$ up to a bounded function
$\varphi(x,y,z,\xi),\ |\varphi|\leq 1$:
\begin{equation}
u=\phi.\varphi,\ \ p=\pm\phi.\sqrt{1-\varphi^2}.               
\end{equation}
Reversely,
\begin{equation} \phi=\pm\sqrt{u^2+p^2},\ \
\varphi=\frac{u}{\pm\sqrt{u^2+p^2}}.        
\end{equation}

The above relations show that instead of $u$ and $p$ we can work with $\phi$
and $\varphi$. Equations (7)-(10) require only $\phi$ to be running wave
along $z$ (in this coordinate system), and $\varphi$ to be bounded function.
In all other respects these two smooth functions are {\it arbitrary}. Hence,
they may be chosen {\it finite} with respect to the spatial coordinates
$(x,y,z)$. Hence, {\it the nonlinear equations} (7)-(10) {\it allow} $(3+1)$
{\it soliton-like solutions}. Note that, since $\varphi$ is bounded, it is
sufficient to choose just $\phi$ to be spatially finite.

We are going to show now that the two functions $\phi$ and $\varphi$ have
a certain invariant sense and can be introduced in a coordinate free way.

First, let's denote by $\alpha$ the invariant
\[
\alpha=\frac{1}{\sqrt{\frac{{\mathbf E}^2+{\mathbf B}^2}{2}}}.
\]
Since ${\mathbf E}.{\mathbf B}=0$ and ${\bf E}^2={\bf B}^2$ we have
$\alpha=|{\bf E}|^{-1}=|{\bf B}|^{-1}$, and the local frame
\[
\chi=\Bigl(\alpha{\mathbf E}, -\alpha\varepsilon {\mathbf B},
-\alpha^2\varepsilon {\mathbf E}\times {\mathbf B}\Bigr)
\]
is orthonormal. This  frame is defined at every point where the field is
different from zero. At every point we have also the frame of unit
vectors generated by the coordiate system chosen:
$$
\chi^o=\Bigl[{\mathbf e_x}=\frac{\partial}{\partial x}=(1,0,0),
\ {\mathbf e_y}=\frac{\partial}{\partial y}=(0,1,0),
\ {\mathbf e_z}=\frac{\partial}{\partial z}=(0,0,1)\Bigr].
$$
We represent now the frame vectors of $\chi$ through the frame vectors of
$\chi^o$, and obtain the matrix
\[
{\cal A}=\left\|\matrix{
\frac{u}{\sqrt{u^2+p^2}} &\frac{-p}{\sqrt{u^2+p^2}} &0 \cr
\frac{p}{\sqrt{u^2+p^2}} &\frac{u}{\sqrt{u^2+p^2}}  &0 \cr
0                        & 0                        &1
\cr}\right\|.
\]
This matrix has three invariants: $I_1=tr{\cal
A}$, $I_2$={\it the sum of all principal minors of second order},
$I_3=det{\cal A}$. We find
\[
I_1=I_2=\frac{2u}{\sqrt{u^2+p^2}}+1;\ \ I_3=det{\cal A}=1.
\]
Clearly, $\frac12(I_1-1)\leq 1$.

Hence, we can define $\phi$ and $\varphi$ through $\alpha$ and these
invariants in the following way:
\begin{equation}
\phi=\pm\sqrt{\alpha^{-2}I_3({\cal A})},\ \              
\varphi=\frac12\left(tr{\cal A}-1\right).
\end{equation}
\vskip 0.5cm
{\bf Definition 2}. The functions $\phi$ and $\varphi$, defined by (23) will
be called {\it amplitude function} and {\it phase function} of the
corresponding nonlinear solution, respectively. The function
$arccos(\varphi)$ will be called {\it phase} of the solution.
\vskip 0.5cm
For $\vec{\cal F}$ and $\vec{\cal M}$ we obtain
\[
\vec{\cal F}=\Bigl[\varepsilon (p_\xi-\varepsilon p_z),
-\varepsilon (u_\xi-\varepsilon u_z), 0\Bigr],
\]
\[
\vec{\cal M}=\Bigl[-(u_\xi-\varepsilon u_z),-(p_\xi-\varepsilon p_z),0\Bigr].
\]
We shall express the scale factor $L$ through $\phi$ and $\varphi$.
We obtain
\[
|\vec{\cal F}|=|\vec{\cal M}|=\frac{|\phi||\varphi_\xi-\varepsilon \varphi_z|}
{\sqrt{1-\varphi^2}}.
\]
Therefore, since $|{\mathbf E}|=|{\mathbf B}|=|\phi|$, the scale factor
$L$ is obtained as function of $\varphi$ and its first derivatives only,
\begin{equation}
L=\frac{|{\mathbf E}|}{|\vec{\cal F}|}= \frac{|{\mathbf
B}|}{|\vec{\cal M}|}= \frac{\sqrt{1-\varphi^2}}{|\varphi_\xi-\varepsilon
\varphi_z|}.           
\end{equation}

Now we shall separate a subclass of nonlinear solutions, called {\it almost
photon-like}, through  the following conditions on $\varphi$ and $L$:
\begin{equation}
u\frac{\partial \varphi}{\partial x}+
p\frac{\partial \varphi}{\partial y}=0,\ \
p\frac{\partial \varphi}{\partial x}-
u\frac{\partial \varphi}{\partial y}=0,\ \                       
(u^2+p^2)\frac{\partial \varphi}{\partial z}=0;\ \
\frac{\partial L}{\partial \xi}=0.
\end{equation}
The {\it invariant sense} of the first three equations of (25) is that
{\it the phase function $\varphi$ is a first integral of the vector fields}
${\mathbf E}, {\mathbf B}, {\mathbf E}\times {\mathbf B}$. From  the third
equation of (25) is clearly seen that in this coordinate system $\varphi$
does not depend on $z$.  The first two equations of (25), considered as an
algebraic linear homogeneous system with respect to the two derivatives,
yield $\varphi_x=\varphi_y=0$ because the corresponding determinant is always
nonzero: $u^2+p^2\neq 0$.  Hence, $\varphi$ may depend only on $\xi$.  In
view of (24) and $\varphi_z=0$
the fourth equation $\frac{{\partial L}}{{\partial \xi}}=0$
means $L= constant$.  Hence, relation (24) turns into equation
for $\varphi$:

\begin{equation}
L=\frac{\sqrt{1-\varphi^2}}{|\varphi_\xi|}\ \ \rightarrow\ \
\frac{{\partial \varphi}}{{\partial \xi}}=                     
\mp \frac1L\sqrt{1-\varphi^2}.
\end{equation}
The obvious solution is (both signs on the right are admissible):
\begin{equation}
\varphi={\rm cos}\left(\kappa\frac{\xi}{L}+\beta_o\right)
={\rm cos}\left(\kappa\frac{c}{L} t+\beta_o\right),               
\end{equation}
where $\kappa=\pm 1,\ \beta_o=const$. Since $\varphi$ is a periodic function
with respect to $t$, then $c/L$ has the physical interpretation of {\it
frequency}, and this frequency has nothing to do with the concept of
frequency in classical electrodynamics since it is defined by $L$, and $L$ is
not defined in Maxwell's theory.

It is clearly seen the consistent translational-rotational behavior of
the solution obtained: the electric and magnetic vectors
\[
{\mathbf E}=\Biggl[\phi\,{\rm cos}\left(\kappa\frac{t}{T} +\beta_o\right),
\phi\,{\rm sin}\left(\kappa\frac{t}{T}+\beta_o\right), 0\Biggr],
\]
\[
{\mathbf B}=\Biggl[\varepsilon\phi\,
{\rm sin}\left(\kappa\frac{t}{T}+\beta_o\right),
-\varepsilon\phi\,{\rm cos}\left(\kappa\frac{t}{T}+\beta_o\right), 0\Biggr]
\]
run along $z$: $\phi=\phi(x,y,\xi\pm z)$, and
rotate (left or right: $\kappa=\pm1$) with the frequency \linebreak
$\nu=(c/L)=1/T$.

In order to separate the photon-like solutions we recall that the photon's
characteristic quantity is its integral intrinsic angular momentum, or spin,
being equal to the Planck's constant $h$. Namely $h$ represents
quantitatively in a unified manner the rotational and translational aspects
of its dynamical nature: for all photons the product $WT$ has the same
value $h$, although $W$ and $T$ may be different for the different photons.
That's why every photon should be able to determine its own scale factor
$L=const$ in order to have a {\it cosine} periodic phase function and
to obey the Planck's law:  $h=WT=WL/c$. The photon's intrinsic periodic
process demonstrates itself in our approach through the (left or right)
rotation of the pair (${\mathbf E},{\mathbf B}$). Since these two vectors
are orthogonal to each other and with equal modules: $|{\mathbf E}|=|{\mathbf
B|}$, the basic local quantity appears to be the area of the square defined
by ${\mathbf E}$ and ${\mathbf B}$ at every point, and this area is equal to
$|{\mathbf E}\times {\mathbf B}|$. During one period $T$ this square performs
a full rotation around the direction of propagation and this gives the
local action $|{\mathbf E}\times {\mathbf B}|.T$. In order to obtain the
integral $T$-action of the solution we have to sum up all these local actions.

The above described idea is easily represented mathematically.
In fact, let $u$, $p$ and $u^2+p^2$ be spatially finite functions. Then the
integral energy
\[
W=\frac{1}{4\pi}\int_{{\cal R}^3}(u^2+p^2)dxdydz\ <\infty
\]
is finite. For every {\it almost photon-like} solution we define the
{\it local spin vector} ${\mathbf S}$ by
\begin{equation}
{\mathbf S}=L^2\frac{\vec{\cal F}\times \vec{\cal M}}{4\pi}=
\frac{{\mathbf E}\times{\mathbf B}}{4\pi}
, \ L=const,\ 0<L<\infty.   
\end{equation}
Now, the {\it integral intrinsic action}, or {\it integral spin} ${\cal S}$
of the solution, is defined by
\begin{equation}
{\cal S}=\int_{[0,T]}\int_{{\cal R}^3}{|{\mathbf S}|}dxdydzdt.   
\end{equation}
We obtain
\begin{equation}
{\cal S}=WT.                                                      
\end{equation}
We note once again that this approach works because $W=const<\infty,\
0<L=const<\infty$ and the solution is soliton-like, i.e. it is finite, it has
periodic dynamical structure and is time-stable. Clearly, no solution
of Maxwell's equations (1)-(2) in the whole space has all these properties.

\vskip 0.5cm
{\bf Definition 3}. A nonlinear solution will be called
{\it photon-like} if it is spatially finite, if it satisfies conditions (25)
and if its integral spin ${\cal S}$ is equal to the Planck constant $h$:\
${\cal S}=h$.

\vskip 0.5cm
Finally, we consider briefly the problem of {\it interference} of
photon-like solutions: if we have two photon-like solutions
\[
{\mathbf E}_1=\Biggl[\phi_1(x,y,\xi+\varepsilon_1 z)
{\rm cos}\left(\kappa_1\frac{\xi}{L_1} +\beta_1\right),
\phi_1(x,y,\xi+\varepsilon_1 z) {\rm sin}\left(\kappa_1\frac{\xi}{L_1}+
\beta_1\right), 0\Biggr],
\]
\[
{\mathbf B}_1=\Biggl[\varepsilon_1\phi_1(x,y,\xi+\varepsilon_1 z)
{\rm sin}\left(\kappa_1\frac{\xi}{L_1}
+\beta_1\right)
-\varepsilon_1\phi_1(x,y,\xi+\varepsilon_1 z)
{\rm cos}\left(\kappa_1\frac{\xi}{L_1}+\beta_1\right),0\Biggr],
\]
\[
{\mathbf E}_2=\Biggl[\phi_2(x,y,\xi+\varepsilon_2 z)
{\rm cos}\left(\kappa_2\frac{\xi}{L_2} +\beta_2\right),
\phi_2(x,y,\xi+\varepsilon_2 z)
{\rm sin}\left(\kappa_2\frac{\xi}{L_2}+\beta_2\right), 0\Biggr],
\]
\[
{\mathbf B}_2=\Biggl[\varepsilon_2\phi_2(x,y,\xi+\varepsilon_2 z)
{\rm sin}\left(\kappa_2\frac{\xi}{L_2}
+\beta_2\right)
-\varepsilon_2\phi_2(x,y,\xi+\varepsilon_2 z)
{\rm cos}\left(\kappa_2\frac{\xi}{L_2}+\beta_2\right),0\Biggr]
\]
we ask: under what conditions their sum $({\mathbf E}_1+{\mathbf E}_2,
{\mathbf B}_1+ {\mathbf B}_2)$ will be again a nonlinear solution? Having
done the corresponding elementary computations, we come to the
following important conclusion: if
\begin{equation}
\varepsilon_1=\varepsilon_2,\ \kappa_1=\kappa_2,\ L_1=L_2         
\end{equation}
then the sum $({\mathbf E}_1+{\mathbf E}_2,
{\mathbf B}_1+ {\mathbf B}_2)$ is again a nonlinear solution.

These relations (31) coincide with  the well known from CED {\it
coherence} conditions.  They say that the two photon-like solutions will
interfere, i.e. their sum will be again a solution, if:  \vskip 0.5cm 1.
They propagate along the same direction:  $\varepsilon_1=\varepsilon_2$,

2. They have the same polarization: $\kappa_1=\kappa_2$,

3. They have the same frequency: $\nu_1=\nu_2$, i.e. the same scale factors.
\vskip 0.5cm
Recall that in CED these coherence conditions
do not follow directly from the theory as necessary conditions, e.g.,
condition 3 requires some time averaging. Moreover, CED
is a linear theory and the sum of {\it any} two or more solutions is again a
solution, for example, the sum "plane wave + a spherically symmetric field"
is again a solution but no interference features are available.

In EED, which is a non-linear theory and there is no superposition principle
in general, the interference of photon-like solutions is a {\it very special
case} and it is a remarkable result that the experimentally found coherence
conditions (31) appear from the nonlinear equations as {\it necessary}
conditions, otherwise the sum {\it will not} be a solution.

The computation shows that the energy density $w$ of the
sum-solution is given by
\begin{equation}
w=\frac{1}{4\pi}\left[(\phi_1)^2+(\phi_2)^2+
2\phi_1\phi_2{\rm cos}(\beta_1-\beta_2)\right],   
\end{equation}
and this relation (32) allows to talk about interference instead of
superposition: in our approach the allowed superposition of photon-like
solutions leads always to interference.

Of course, from the non-linear point of view, these interference phenomena
are of some interest only for those soliton-like solutions which at a given
moment occupy intersecting regions, otherwise the interference term
$2\phi_1\phi_2{\rm cos}(\beta_1-\beta_2)$ in (32) is equal to zero. Since the
two summonds follow the same direction of motion as a whole, then the
sum-solution will be a time-stable solution, but it {\it will not be}
photon-like one.

\section{Conclusion}
This paper presents a non-relativistic formulation of an extension of the
pure field Maxwell equations. The main purpose of the extension is to give a
mathematical description of the viewpoint that photons are extended but
finite objects and that their existence is based on a joint and consistent
translational-rotational internal dynamics: a straight-line (translational)
motion as a whole with the velocity of light and a rotation of the mutually
orthogonal and perpendicular to the translation electric and magnetic vectors.
The general mathematical concept of $(3+1)$-solitary wave, or soliton, was
found to be the most adequate one to this physical notion.  Our nonlinear
equations (7)-(10) for the couple of vector fields $({\bf E},{\bf B})$
realize {\it directly} the idea for local energy-momentum conservation, i.e.
the idea that the free field dynamics (or time evolution) is determined by a
definite and permanent intra-field local energy-momentum redistribution.
Compare to Maxwell's approach, this is a new look on the field's evolution
defining equations, it carries the idea of the second Newton's law in
mechanics to continuously distributed in the 3-space physical systems
(electromagnetic fields). It is also a new moment that the field has more
than one potential abilities to exchange energy-momentum with other physical
systems.  This nonlinear approach turned out to be successful in view of the
existence of appropriate photon-like solutions.  Every photon-like solution
has finite integral energy $W$, has its own scale factor $L=const$, phase
function $\varphi$ of $cosine$ type and corresponding to $L$ and $\varphi$
frequency $\nu=c/L$, or period $T=L/c$.  Every photon-like solution carries
intrinsic angular momentum of integral value equal to the Planck's constant
$h=WT$, which is the famous Planck's formula.  The equations (7)-(10) and the
additional conditions (25) do not determine the spatial structure of the
solution and it should be so, because it is hardly believable that all
photons must have the same shape, structure and extension.  Moreover, if the
amplitude function $\phi$ consists of many non-overlaping appropriate bumps,
we obtain a mathematical image of a number of coherent photons, i.e. a
many-bump photon-like solution, and the corresponding integral spin momentum
will be $nh$, where $n$ gives the number of bumps. Clearly, large enough
parts of the classical plane wave can be considerd as macro-pictures of such
a many-bump photon-like solution.  Nonlinear solutions with intrinsic angular
momentum different from $h$, or $nh$, where $n=1,2,. . . $, are also, in
principle, allowed. The remarkable limited superposition principle permits
interference in the frame of photon-like solutions only if they are coherent
in the sense of (31).

The existence of localized photon-like solutions in the pure field case
suggests to make the corresponding extension of Maxwell's equations in
presence of external fields (media) and to look for (3+1)-localized solutions
with non-zero proper mass. Such an extension in relativistic terms was made
and published [2], where a large family of variously shaped $(3+1)$
soliton-like solutions with non-zero mass and well defined conserved
quantities is also given.

\vskip 0.5cm
We would like to acknowledge the substantial support of the
Bulgarian National Fund "Science Research" through Project $\phi-718$ in
doing this research.

\newpage
REFERENCES
\vskip 1 cm

[1]. {\bf S.Donev, M.Tashkova}, Proc.R.Soc.of Lond., A 443, (301), 1993.
\vskip 0.5cm
[2]. {\bf S.Donev, M.Tashkova},  Proc.R.Soc. of Lond., A 450, (281), 1995.
\vskip 0.5cm
[3]. {\bf S.Donev, M.Tashkova}, Annales de la Fondation Louis de Broglie,
vol.23, No.No.2,3 (1998).
\vskip 0.5cm
[4]. {\bf S.J.Farlow}, "Partial Differential Equations for Scientists and
Engineers", John Wiley \& Sons, Inc., 1982.
\vskip 0.5cm
[5]. {\bf G.N.Lewis}, Nature {\bf 118}, 874, 1926.
\vskip 0.5cm
[6]. {\bf A.Einstein}, Sobranie Nauchnih Trudov, vols.2,3, Nauka,
Moskva, 1966.
\vskip 0.5cm
[7]. {\bf M.Planck}, J.Franklin Institute, 1927 (July), p.13.
\vskip 0.5cm
[8]. {\bf J.J.Thomson}, Philos.Mag.Ser. 6, 48, 737 (1924), and 50, 1181
(1925), and Nature, vol.137, 23 (1936); {\bf N.Rashevsky}, Philos.Mag. Ser.7,
4, 459 (1927); {\bf W.Honig}, Found.Phys. 4, 367 (1974); {\bf G.Hunter,
R.Wadlinger}, Phys.Essays, vol.2,158 (1989).
\vskip 0.5cm
[9]. {\bf A.Einstein}, J.Franklin Institute, 221 (349-382), 1936.
\vskip 0.5cm
[10]. {\bf J.D.Jackson}, {\it CLASSICAL ELECTRODYNAMICS}, John Wiley and
Sons, Inc., New York-London, 1962.
\vskip 0.5cm

\end{document}